\documentclass[
 reprint,
 amsmath,amssymb,
 aps,
]{revtex4-2}

\usepackage{graphicx}
\usepackage{dcolumn}
\usepackage{bm}
\usepackage{xcolor}
\usepackage[normalem]{ulem} 

\begin{document}


\title{Scaling of percolation transitions on Erd\"os-R\'enyi networks under centrality-based attacks}

\author{Nahuel Almeira}
 \altaffiliation{}
 \email{nalmeira@famaf.unc.edu.ar}%
\author{Orlando Vito Billoni}%
 \email{billoni@famaf.unc.edu.ar}
\affiliation{%
 Facultad de Matem\'atica, Astronom\'{\i}a, F\'{\i}sica y Computaci\'on, Universidad Nacional de C\'ordoba\\
Ciudad Universitaria, 5000 C\'ordoba, Argentina
 }
\affiliation{%
Instituto de F\'{\i}sica Enrique Gaviola (IFEG-CONICET)\\
Ciudad Universitaria, 5000 C\'ordoba, Argentina
}%

\author{Juan Ignacio Perotti}
\affiliation{%
Instituto de F\'{\i}sica Enrique Gaviola (IFEG-CONICET)\\
Ciudad Universitaria, 5000 C\'ordoba, Argentina
}%

\date{\today}

\begin{abstract}
The study of network robustness focuses on the way the overall functionality of a network is affected as some of its constituent parts fail. 
Failures can occur at random or be part of an intentional attack and, in general, networks behave differently against different removal strategies. Although much effort has been put on this topic, there is no unified framework to study the problem. 
While random failures have been mostly studied under percolation theory, targeted attacks have been recently restated in terms of network dismantling. In this work, we link these two approaches by performing a finite-size scaling analysis to four dismantling strategies over Erd\"{o}s-R\'enyi networks: initial and recalculated high degree removal and initial and recalculated high betweenness removal. We find that the critical exponents associated with the initial attacks are consistent with the ones corresponding to random percolation, while the recalculated attacks are likely to belong to different universality classes. In particular, recalculated betweenness produces a very abrupt transition with a hump in the cluster size distribution near the critical point, resembling some explosive percolation processes. 
\end{abstract}

\maketitle

\section{\label{sec:Introduction}Introduction}

The study of percolation in complex networks is a current research topic that has both theoretical inquires~\cite{DSouza2015AnomalousPercolation,Lee2018RecentNetworks} and practical applications. Percolation transitions are observed in many biological, social and technological complex networks~\cite{zamponi2018mitochondrial,zeng2019switch} 
and are connected to the problem of resilience to damage~\cite{Tian2017ArticulationNetworks,DaCunha2015FastAttacks,daCunha2017PerformanceNetworks,Shekhtman2015ResilienceNetworks} and therefore to the functionality of the systems associated with the networks.
Also, theoretical tools devised for the analysis of percolation have been used in the study of disease spreading~\cite{Valdez2012TemporalSpreading}, city traffic dynamics~\cite{Zeng2017SwitchDynamics}, and the structural characterization of networks~\cite{Allard2018PercolationNetworks}, among others. 
In particular, mathematical models for percolation processes on inter-dependent graphs where developed~\cite{Buldyrev2010} to capture salient features of random failures of systems such as power grids.

Failures are usually modeled as random deletions of nodes or links, while in attacks influential nodes or links are removed according to a rank of specific characteristics, trying to produce the greatest damage to the network. The effectiveness of the attack strategy depends on the topological features of the network as well as on the type of attack. For this reason, several network architectures were studied under different strategies to evaluate both the robustness of networks and the effectiveness of attacks~\cite{Holme2002AttackNetworks, Iyer2013AttackNetworks, Wandelt2018ANetwork}. In the pioneering work of Holme et al. \cite{Holme2002AttackNetworks}, the effect of different centrality edge- and node-based attacks was studied in several synthetic and real-world networks. Later, Iyer et al. extended these results by studying new centrality measures and network  models for the case of node-based attacks. In a recent study performed by Wandelt et al.~\cite{Wandelt2018ANetwork}, an extensive benchmark of synthetic and real-world networks was analyzed using different dismantling strategies
providing the most extensive comparative analysis on network robustness up to now.

It is widely known that scale-free networks are fragile against centrality targeted attacks \cite{Callaway2000NetworkGraphs,Albert2000,Cohen2000BreakdownAttack,Holme2002AttackNetworks} but robust against random failures~\cite{Cohen2000BreakdownAttack}. 
On the other hand, networks with homogeneous degree distribution, such as Erd\H{o}s-R\'enyi (ER) networks, are expected to be robust under targeted attacks. 
In particular, they have been proved to be robust against degree-based attacks~\cite{Crucitti2004}. However, some networks that are fragile to targeted attacks despite having homogeneous degree distributions. One such example is the US power grid, which exhibits a significant connectivity loss when nodes with high load are deleted~\cite{Albert2004StructuralGrid}. Another example is the Watts-Strogatz model of small-world homogeneous networks since they have been proved to be particularly fragile in a cascading failure scenario.
Xia et al. ~\cite{Xia2010CascadingNetworks} attributed the fragility of these networks to their heterogeneous betweenness distribution.
Attacks based on betweenness are among the most efficient ways to dismantling a network~\cite{Holme2002AttackNetworks,daCunha2017PerformanceNetworks,Iyer2013AttackNetworks, Wandelt2018ANetwork} and are particularly effective in networks having a heterogeneous betweenness distribution. However, in Erd\H{o}s-R\'enyi networks, where both degree and betweenness distributions are homogeneous~\cite{Xia2010CascadingNetworks, Kornbluth2018NetworkAttacks}, a betweenness-based attack is not expected to outperform other targeted attacks. As we will show in this article, this is not the case. In particular, the recalculated version of the betweenness-based attack on nodes is particularly effective to destroy ER networks, with a performance comparable to the most efficient methods to dismantle networks~\cite{Braunstein2016NetworkDismantling,Morone2015InfluencePercolation}.
 
In this work, we study percolation processes on ER networks under different attack strategies using finite-size scaling analysis to assess the nature of the transition towards the fragmented phase.
Our results show that the choice of the attack strategy can change the properties of the transition. In particular, the transition produced by the recalculated betweenness-based attack is sharper than for the rest of the attacks, deviating significantly from the random percolation universality class. 
Given the steep variation of the order parameter near the transition, we consider the process as a case of ``explosive percolation''~\cite{Achlioptas2009ExplosiveIn,DaCosta2014SolutionExponents,DSouza2019ExplosiveNetworks}. 
The results of the finite-size scaling analysis are consistent with a continuous phase transition, but we cannot ensure that this result holds in the infinite-size limit.

\subsection{Attack strategies}

In centrality-based attacks, nodes are sorted in decreasing order according to a centrality measure. Then, they are sequentially removed according to that list (ties, if any, are usually broken randomly). There is an extensive list of centrality measures that have been tested in multiple networks (see, for example, \cite{Iyer2013AttackNetworks}). Some of the most popular are degree, betweenness \cite{Freeman1977ABetweenness}, closeness, eigenvector and collective influence \cite{Morone2015InfluencePercolation}. In general, when a node is removed, the centrality values of the remaining nodes change. Thus, the attack can be improved by recomputing the list after each removal step. If the centrality measure uses only local information, like degree or collective influence, only a fraction of nodes will eventually change, so the original ordering of the nodes may remain the same after several steps. On the other hand, measures like betweenness or eigenvector centrality use global information, so even the deletion of a single node can potentially change the ordering in a significant way. Given that the \emph{recalculated} version of an attack uses more updated information of the network, it is in general more efficient than its \emph{initial} counterpart \cite{Iyer2013AttackNetworks}.

In this work, we will focus on both the initial and recalculated versions of the attacks based on two centrality measures: degree and betweenness.
 The degree of a node, defined as the number of neighbors the node has, is the most intuitive centrality measure and one of the most studied in the literature. It is easily interpreted in terms of network connectivity and it has the advantage of being a local measure, which makes it suitable for analytical treatment. On the other hand, betweenness centrality is a global measure and is defined in the following way. Let $\sigma(s, t)$ be the number of shortest paths connecting nodes $s$ and $t$ and let $\sigma_i(s, t)$ the number of such paths going through node $i$. Then, the betweenness centrality of node $i$ is
\begin{equation}
    b_i = \sum_{s\neq t} \dfrac{\sigma_i(s, t)}{\sigma(s,t)},
\end{equation}
where we adopt the convention that $\sigma_i(s, t)/\sigma(s,t) = 0$ if both $\sigma_i(s, t)$ and $\sigma(s, t)$ are zero.

Betweenness can be thought of as the amount of load a node must support when there is some kind of flux on the network. Nodes with higher betweenness articulate different groups of nodes and their importance is more related to the communicability of the network.
In particular, it is easy to check that nodes with degree lower than two have betweenness equal to zero.
Being a global measure, it is hard to compute this centrality. 
The most efficient algorithm so far known was proposed by Brandes, \emph{et al.} in \cite{Brandes2001ACentrality} and runs like $\mathcal{O}(NM)$, where $N$ and $M$ are the number of nodes and links in the network, respectively. The main reason for considering this measure is that it has been reported as the most efficient attack strategy for many networks, including both synthetic and real-world networks \cite{Iyer2013AttackNetworks, Wandelt2018ANetwork}.

\subsection{Percolation}\label{subsec:Percolation}

Site percolation in complex networks can be stated by considering that each node of the network can be either \emph{occupied}, with probability $p$, or unoccupied, with probability $1-p$. Only occupied nodes can be connected, thus links connecting at least one unoccupied node are also considered unoccupied. If $p=0$, the network is empty and if $p=1$, the original network is recovered. When the occupation probability is small, occupied nodes belong to different small-sized components, but above a critical value $p=p_c$, one of the components acquires an extensive size. At this point, it is said that the system percolates. The extensive component is known as the \emph{giant connected component} (GCC) and the critical point is referred to as the \emph{percolation threshold}.

Let $N$ be the size of the network and $N_1$ the size of the GCC. In the thermodynamic limit $N\rightarrow \infty$, percolation theory states that the relative size $S_1 = N_1/N$ follows the critical behavior
\begin{equation}
    S_1=
    \begin{cases}
    0 &  p <  p_c, \\
    a(p-p_c)^\beta &  p \ge p_c,
    \end{cases}
\end{equation}
where $a$ is a proportionality constant and $\beta>0$ is the critical exponent associated with $S_1$. The transition between the percolated and non-percolated state has been widely studied in statistical physics, and it has been shown to exhibit a continuous transition in many different network models. In this framework, $S_1$ is considered the order parameter of the transition. 

As it occurs in continuous transitions, other measures also manifest a critical behavior near the percolation threshold. One such measure is the average cluster size, which plays the role of susceptibility and is computed as 
\begin{equation}
    \langle s \rangle = \frac{\sum_{s}^{'} s^2 n_s(p)}{\sum_s^{'} s n_s(p)} 
\end{equation}
where $n_s(p)$ is the number of clusters of size $s$ per node and the primed sum excludes the GCC. At the critical point, $\langle s \rangle$ diverges in the thermodynamic limit as $\langle s \rangle  \sim (p-p_c)^{-\gamma}$, with $\gamma>0$. Also, $n_s(p)$ has its own critical behavior and close to $p_c$ it becomes very heterogeneous, being well described by the expression
\begin{equation} \label{eq:ns}
n_s(p) \sim s^{-\tau} e^{-s/s^*}.
\end{equation}
Here $s^*$ represents the characteristic cluster size, which scales as $s^* \sim |p-p_c|^{-1/\sigma}$. Then, at $p=p_c$ the number of clusters of size $s$ follows a power-law $n_s(p) \sim s^{-\tau}$. Finally, the correlation length $\xi$, defined as the geometrical length of a typical cluster, scales as $\xi \sim (p-p_c)^{-\nu}$, where $\nu>0$~\cite{StaufferBook}.

The theory of critical phenomena states that continuous transitions can be fully characterized by its critical exponents. If the same exponents are shared between two systems, they belong to the same universality class. In percolation only two exponents are independent, and the others can be derived using different scaling relations. For example, the exponent associated with the cluster size distribution can be obtained as \cite{StaufferBook}
\begin{equation}\label{eq:scaling_tau}
    \tau = 2 + \dfrac{\beta}{\gamma + \beta}.
\end{equation}
As $\beta$ and $\gamma$ are both positive, equation \ref{eq:scaling_tau} shows that $\tau \geq 2$. Another useful relation is given by \cite{BrankovBook, Fortunato2011ExplosiveGraphs}
\begin{equation} \label{eq:gamma_beta_nu}
    2\dfrac{\beta}{\bar{\nu}} + \dfrac{\gamma}{\bar{\nu}} = 1,
\end{equation}
where $\bar{\nu} = d\nu$ and $d$ is the effective dimension of the network.

 Standard site percolation on Erd\H{o}s-R\'enyi graphs reports the mean-field exponents, with 
$\beta = \gamma =  1$, $\bar{\nu} = 3$, $\sigma = 1/2$, and $\tau = 5/2$ \cite{Albert2002StatisticalNetworks,Boccaletti2016ExplosiveSynchronization}. Also, in uncorrelated networks, the percolation threshold is given by~\cite{Cohen2000ResilienceBreakdowns}
\begin{equation}
    p_c = \dfrac{1}{\kappa-1},
\end{equation}
where $\kappa = \langle k^2\rangle / \langle k \rangle$ is the heterogeneity parameter of the degree distribution.

From a theoretical point of view, standard percolation and node removal are different processes~\cite{Cohen2000BreakdownAttack}. Percolation is an equilibrium reversible process, well described by the equilibrium statistical physics. 
On the other hand, node removal under specific attacks
are irreversible  processes such as the evolving rules that turn out in explosive percolation transitions~\cite{DaCosta2014SolutionExponents}.
Being aware of this, we relate the percolation probability $p$ with a node removal procedure in which a fraction $f=1-p$ of nodes was removed. Using this relation we can apply the tools provided by percolation theory to the attack strategies previously described.

\section{Results}\label{sec:Results}


\subsection{Percolation transition}

\begin{figure}
    \centering
    \includegraphics[scale=0.26]{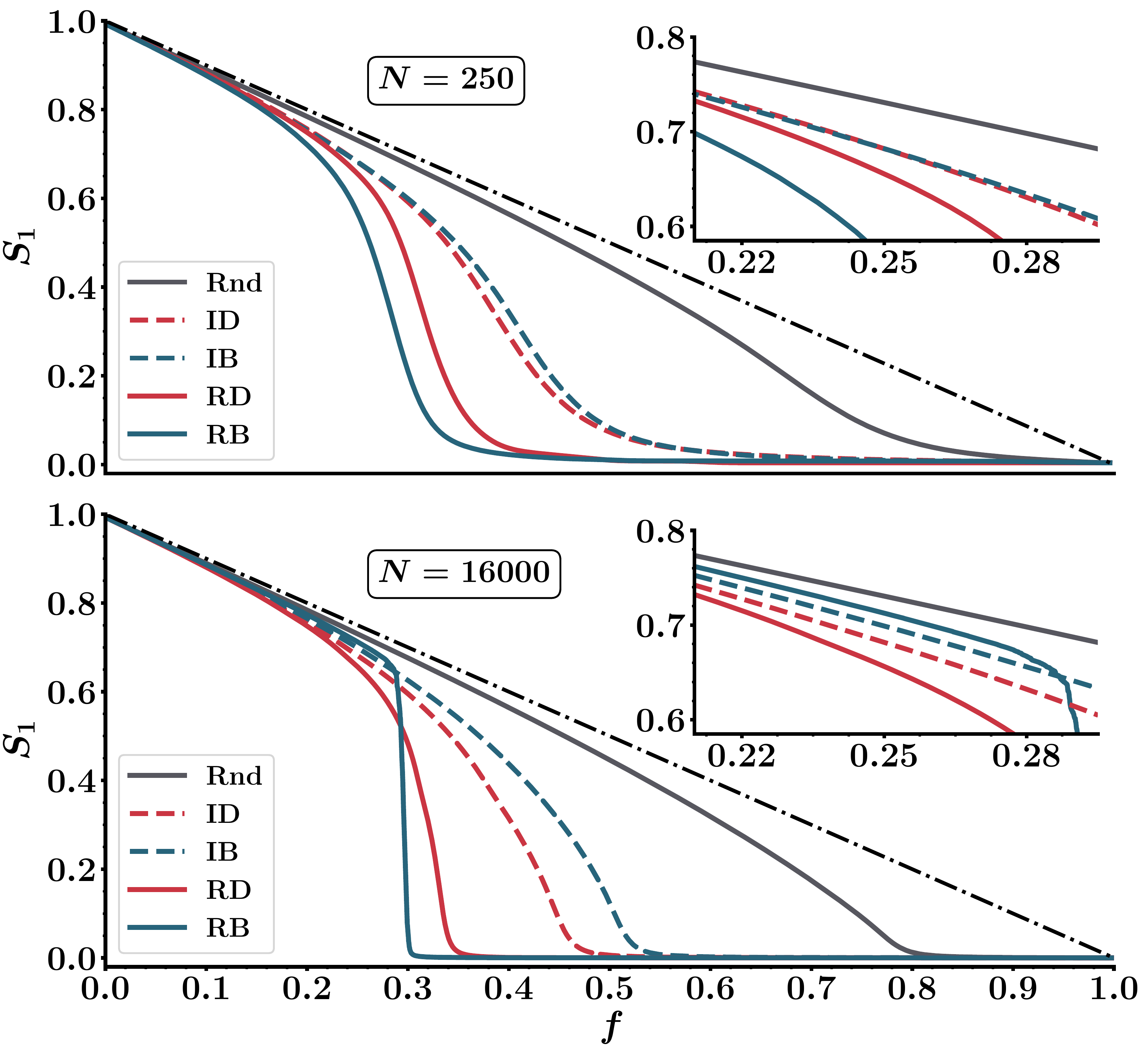}
    \caption{Relative size of the giant component as a function of the fraction of removed nodes, averaged over $100$ realizations, for two ER networks with $\langle k \rangle = 5$ and different sizes. 
    As a reference, the dot-dashed line corresponds to node removal of a fully connected graph. The inset zooms the behavior right before the collapse produced by RB.}
    \label{fig:percolation}
\end{figure}

Figure~\ref{fig:percolation} shows the evolution of the size of the giant component as a function of the fraction of removed nodes $f$ on an ER network with mean degree $\langle k \rangle=5$. Each curve, which is an average taken over $100$ independent networks, corresponds to a different attack, namely recalculated betweenness (RB), recalculated degree (RD), initial betweenness (IB), initial degree (ID) and random removal (Rnd). When a small network is considered (upper panel), it can be seen that ID performs slightly better than IB, in the sense that, for each fraction of nodes removed the network is consistently more fragmented when nodes with high degree are removed.
As it has been previously reported by Iyer, et al. in \cite{Iyer2013AttackNetworks}, the situation reverts when the list of nodes is recalculated after each node removal, with RB outperforming RD. When a bigger network with the same characteristics is attacked (lower panel), all the transitions become sharper. Except perhaps for RB, all the curves seem to be consistent with a continuous percolation transition. The curve for recalculated betweenness, on the other side, exhibits a very abrupt collapse at $f\approx 0.3$, with a very steep slope. Interestingly, for lower values of $f$ this attack performs poorly (see inset), barely outperforming random removal. In fact, at the beginning both attacks (RB and Rnd) do not produce a network fragmentation as can be seen when compared to the attack of a fully connected graph in which the network is reduced one node at a time.


\subsection{Finite-size scaling analysis}

Finite-size scaling analysis is one of the most important tools in the study of  continuous phase transitions and in particular to obtain the associated critical exponents~\cite{Cho2010Finite-sizeTransitions, Fortunato2011ExplosiveGraphs, Zhu2017FiniteTransition}.
According to this theory, the divergence of the correlation length at the critical point implies that every variable of the system becomes scale-independent at this point. For a finite-size system of size $N$, this produces a scaling of the form
\begin{equation}
    X \sim N^{-\omega/\bar{\nu}} F[(f-f_c) N^{1/\bar{\nu}}],
\end{equation}
where $\omega$ is an exponent related to the variable $X$. For $f=f_c$, the variable behaves as $X \sim N^{-\omega/\bar{\nu}}$. This relation holds asymptotically, i.e. in the limit $N \rightarrow \infty$ and $f \rightarrow f_c$, and it can be used to obtain the ratio $\omega/\bar{\nu}$ by computing $X(f_c, N)$ for different system sizes. In addition, the plot of $N^{\omega/\bar{\nu}}X$ as a function of $(f_c-f) N^{1/\bar{\nu}}$ yields to the universal function $F$, which does not depend on $N$, so curves corresponding to different sizes collapse.

\begin{figure}[h]
    \centering
    \includegraphics[scale=0.21]{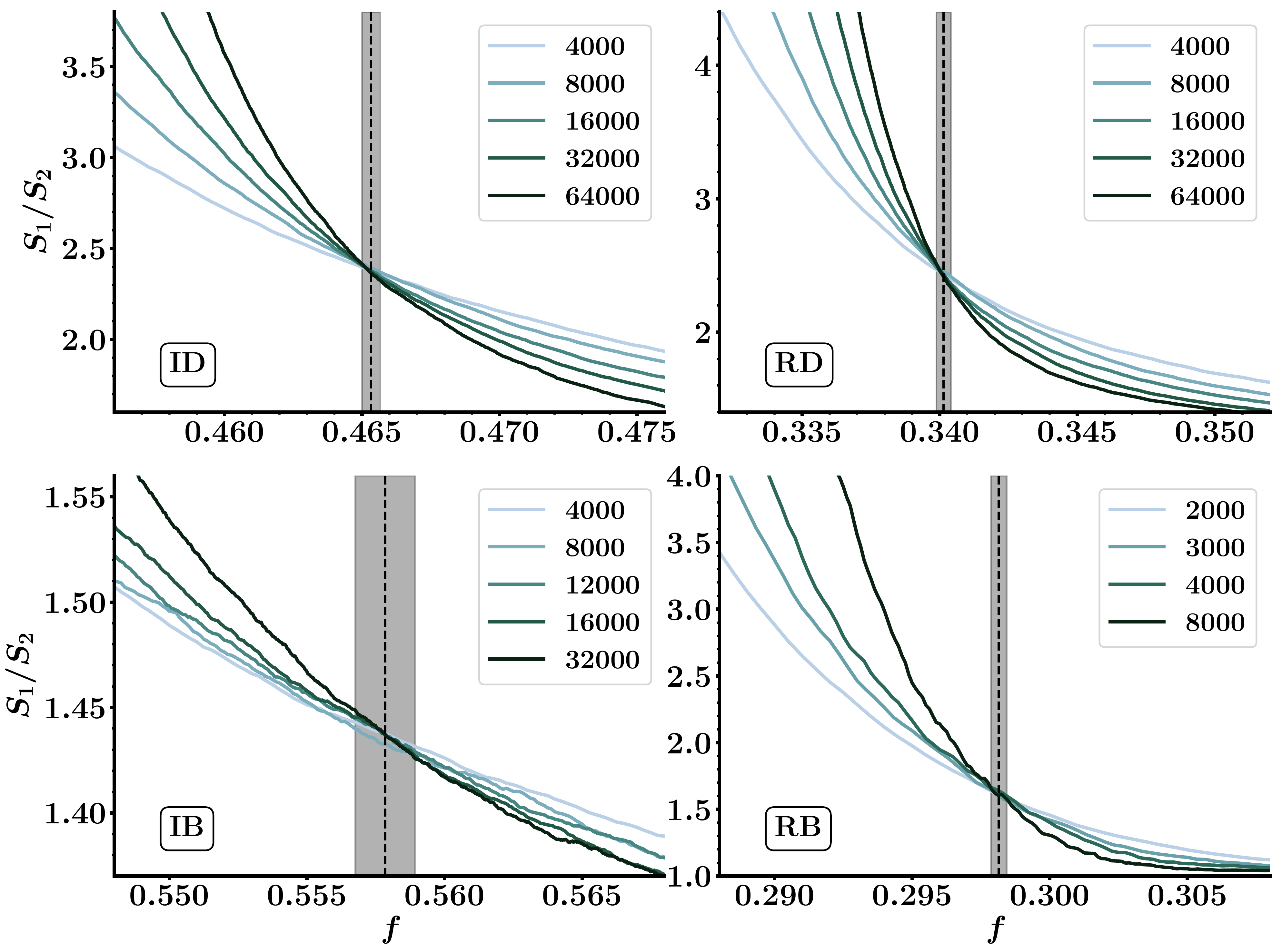}
    \caption{ Determination of the critical point $f_c$ through the crossing-point method for the four attack strategies considered. Each curve represents an average over $2\times 10^4$ independent networks for the attacks ID, RD, and IB and over $10^3$ networks for RB. The vertical line corresponds to the mean of the intersections and the shadow region to the standard deviation. The values obtained were $f_c^{\mathrm{(ID)}} = 0.4652(7) $, $f_c^{\mathrm{(RD)}} = 0.3401(2)$, $f_c^{\mathrm{(IB)}} = 0.558(1)$ and $f_c^{\mathrm{(RB)}} = 0.2984(2)$.}
    \label{fig:crossing_rest}
\end{figure}

In this work, we make use of two scaling relations. The first one is the scaling of the cluster relative sizes, which can be stated as  \cite{Zhu2017FiniteTransition}
\begin{equation}
\label{eq:comp-scaling}
S_i(f,N) \sim N^{-\beta /\bar{\nu}} \bar{S}_i[(f-f_c) N^{1/\bar{\nu}}].
\end{equation}
 Here, the subscript $i=1,2, ...$ indicates the rank of each component, sorted by size in decreasing order. In particular, we will be interested in the order parameter $S_1$ and in the size of the second cluster $S_2N$. The second scaling relation involves the average cluster size and can be stated as
\begin{equation}
\label{eq:suscep-scaling}
\langle s \rangle(f,N)  \sim N^{\gamma / \bar{\nu}} \tilde{S} [(f-f_c) N^{1/\bar{\nu}}].
\end{equation}
The percolation threshold $f_c$ can be determined in several ways (for some of them, see \cite{Radicchi2009ExplosiveNetworks}).
The alternative we used in our work, which we call the crossing-point method, is the following. We first define $S_{ic}(N)\equiv S_i(f_c,N)$. According to Eq. \ref{eq:comp-scaling}, we have
\begin{equation} \label{eq:scaling_SM}
    S_{ic}(N) \sim N^{-\beta /\bar{\nu}} \tilde{S}_i(0).
\end{equation}
Now we take the relative size of the first two components for a given size $N$ and compute the quotient between them. The resulting expression becomes $S_{1c}(N)/S_{2c}(N) \sim \tilde{S}_1(0)/\tilde{S}_2(0)$, which is independent of $N$. This result implies that at the critical point $f=f_c$, the curves of $S_{1}(f, N)/S_{2}(f, N)$ for different sizes should take the same value. The crossing-point method consists of numerically estimate the intersection of these curves. Using averages over $2\times10^4$ independent networks for the attacks ID, RD, and IB and over $10^3$ independent networks for RB, we computed the quotients $S_1(f, N)/S_2(f, N)$ for different sizes and then calculated the values of the intersections for each pair of sizes. The value of $f_c$ that we report is the mean of these intersections, with the standard deviation as the associated uncertainty.
The method, applied to the four attack strategies previously described, is shown in Fig. \ref{fig:crossing_rest}. As it can be seen in the figure, the performance of the method depends on the nature of the attack. For the two degree-based attacks, the intersections occur with a very low variance, so the percolation threshold can be estimated with a very high precision as $f_c^{\mathrm{(ID)}} = 0.4652(7)$ and $f_c^{\mathrm{(RD)}} = 0.3401(2)$. The initial betweenness-based attack, on the other hand, has a lower precision, and the value obtained was $f_c^{\mathrm{(IB)}} = 0.558(1)$. 
The method also performs very well for the recalculated betweenness-based attack, even when the sizes and number of simulations that were used are lower, due to its high computational complexity, giving an estimated percolation threshold of $f_c^{\mathrm{(RB)}} = 0.2984(2)$. 
From a dismantling point of view, the recalculated versions of the  attacks are more effective than their initial counterparts, since they have lower percolation thresholds. 
In particular, the initial version of the betweenness-based attack is a rather poor dismantling strategy, being closer to random node removal than the rest of the attacks. 
On the other hand, the recalculated version of this attack is the most efficient one, performing better or comparable with other state-of-the-art dismantling strategies \cite{Morone2015InfluencePercolation,Braunstein2016NetworkDismantling}. 

%

\begin{figure}[h]
    \centering
    \includegraphics[scale=0.21]{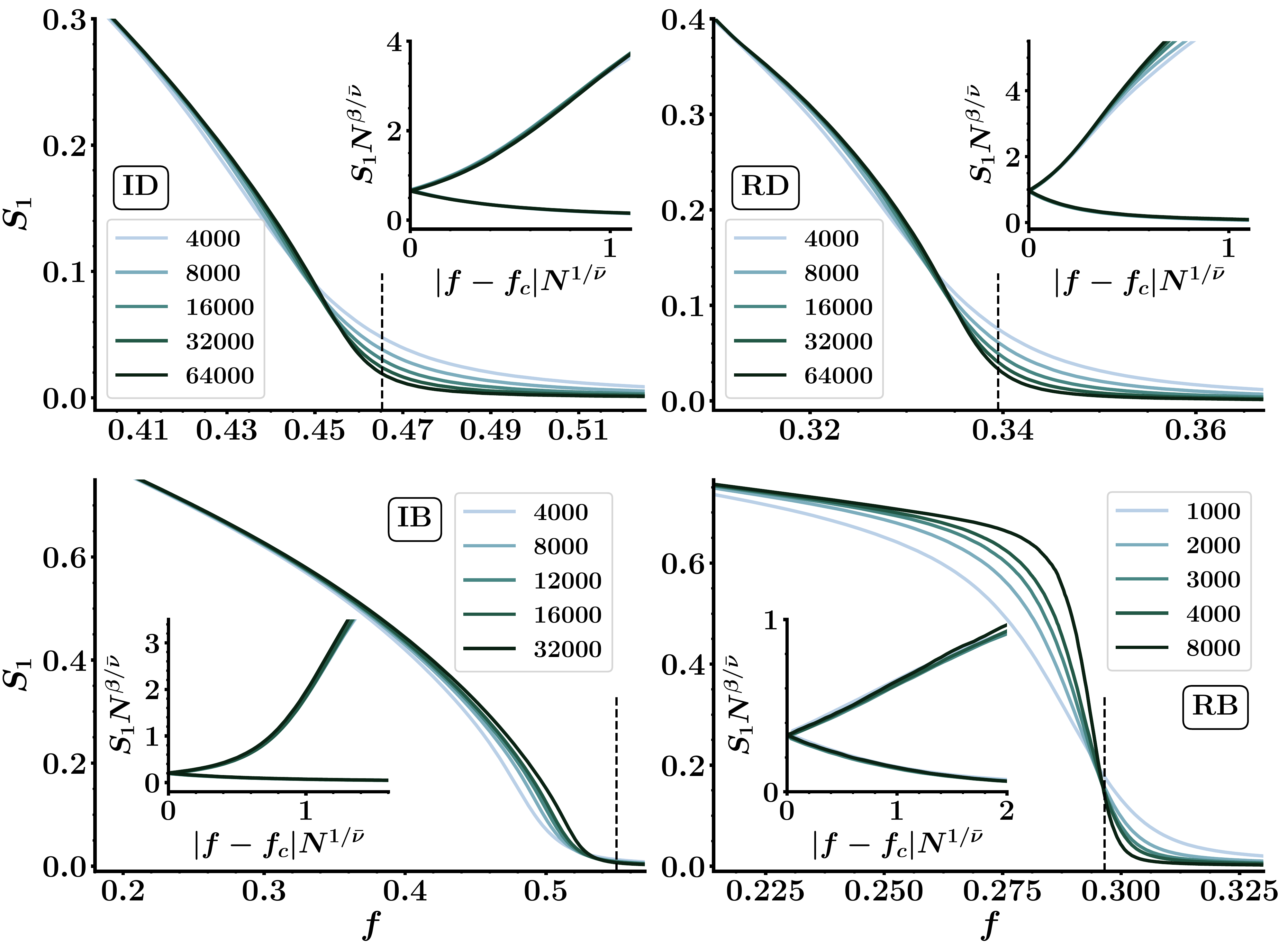}
    \caption{Order parameter $S_1$ as a function of the fraction of nodes removed $f$ in the neighborhood of the critical point. Each panel corresponds to one of the four attack strategies studied. Data is averaged over $2\times 10^4$ realizations for the attacks ID, RD, and IB and over $10^3$ realizations for RB. The dashed vertical lines correspond to the value of the percolation threshold computed using the crossing method (see main text). The insets show the collapse of the curves using the scaling ansatz given by Eq. \ref{eq:comp-scaling}. The values for the percolation thresholds and critical exponents used to perform the scaling are the ones summarized in Table \ref{table}.}
    \label{fig:S1_scaling}
\end{figure}

Once the percolation threshold has been estimated, we proceed to study the order parameter, susceptibility and second cluster size in the vicinity of the transition. The four panels of Fig. \ref{fig:S1_scaling} show the order parameter $S_1$ as a function of the fraction of nodes removed. Each panel corresponds to a different attack and each curve in the main panels corresponds to a different system size. It can be seen that the transitions become sharper as $N$ increases, particularly in the case of RB attack. The curves of each panel can be collapsed on both sides of the transition (see insets) using the scaling relation of Eq. (\ref{eq:comp-scaling}). The exponents $\beta/\bar{\nu}$ and $\bar{\nu}$ used to collapse the curves corresponding to each attack, which were estimated using the methods described below, are compiled in Table \ref{table}.

\begin{table}[h!]
    \begin{center}
        \begin{tabular}{|c|c|c|c|c|c|}
        \hline
         & $f_c$ & $\beta/\bar{\nu}$ & $\gamma/\bar{\nu}$ & $\bar{\nu}$ & $\tau$
        \\
        \hline
        Rnd & $0.8$ & $1/3$ & $1/3$ & $3$ & $2.5$
        \\
        \hline
        ID & $0.4652(7)$ & $0.320(4)$ & $0.354(5)$ & $2.72(5)$ & $2.51(4)$ $[2.47(2)]$
        \\
        \hline
        RD & $ 0.3401(2)$ & $0.307(3)$ & $0.377(7)$ & $2.59(7)$ & $2.37(4)6$ $[2.45(2)]$ 
        \\
        \hline
        IB & $ 0.558(1)$ &  $0.340(4)$ & $0.335(6)$ & $2.8(2)$ &  $2.51(4)$  $[2.50(2)]$
        \\
        \hline
        RB & $0.2984(2)$  & $0.10(2)$ & $0.89(2)$ & $1.50(5)$ & -- $[2.1(2)]$
        \\
        \hline
        \end{tabular}
        \caption{Numerical estimation of the percolation thresholds and critical exponents for the different attacks in ER networks with $\langle k \rangle = 5$ using finite-size scaling. The values of $\tau$ between brackets were computed using equation \ref{eq:scaling_tau}. }
        \label{table}
    \end{center}
\end{table}

For the four attacks, it can be observed that all the curves collapse very well to a master curve in the proximity of the percolation threshold, confirming that the scaling relations of Eq. (\ref{eq:comp-scaling}) holds. 

\begin{figure}[h]
    \centering
    \includegraphics[scale=0.21]{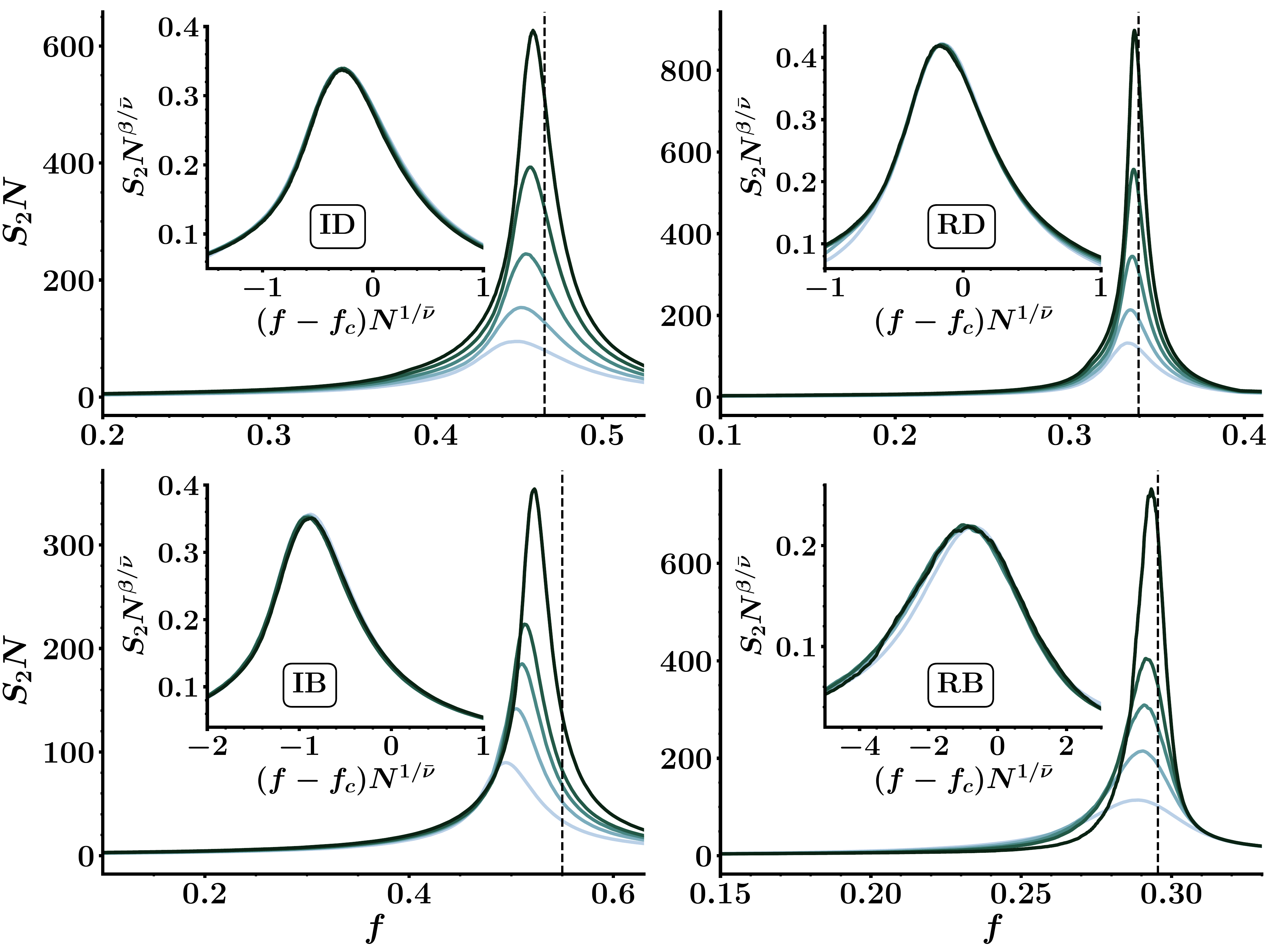}
    \caption{Size of the second cluster $S_2N$ as a function of the fraction of nodes removed $f$ for the four attack strategies studied. As it can be seen, this quantity peaks in the neighborhood of the percolation threshold (dashed vertical line). The color code is the same as in Figure \ref{fig:S1_scaling}. The insets show the collapse of the curves using the scaling ansatz given by Eq. \ref{eq:comp-scaling}, using the parameters given in Table \ref{table}.}
    \label{fig:N2_scaling}
\end{figure}
\begin{figure}[h]
    \centering
    \includegraphics[scale=0.21]{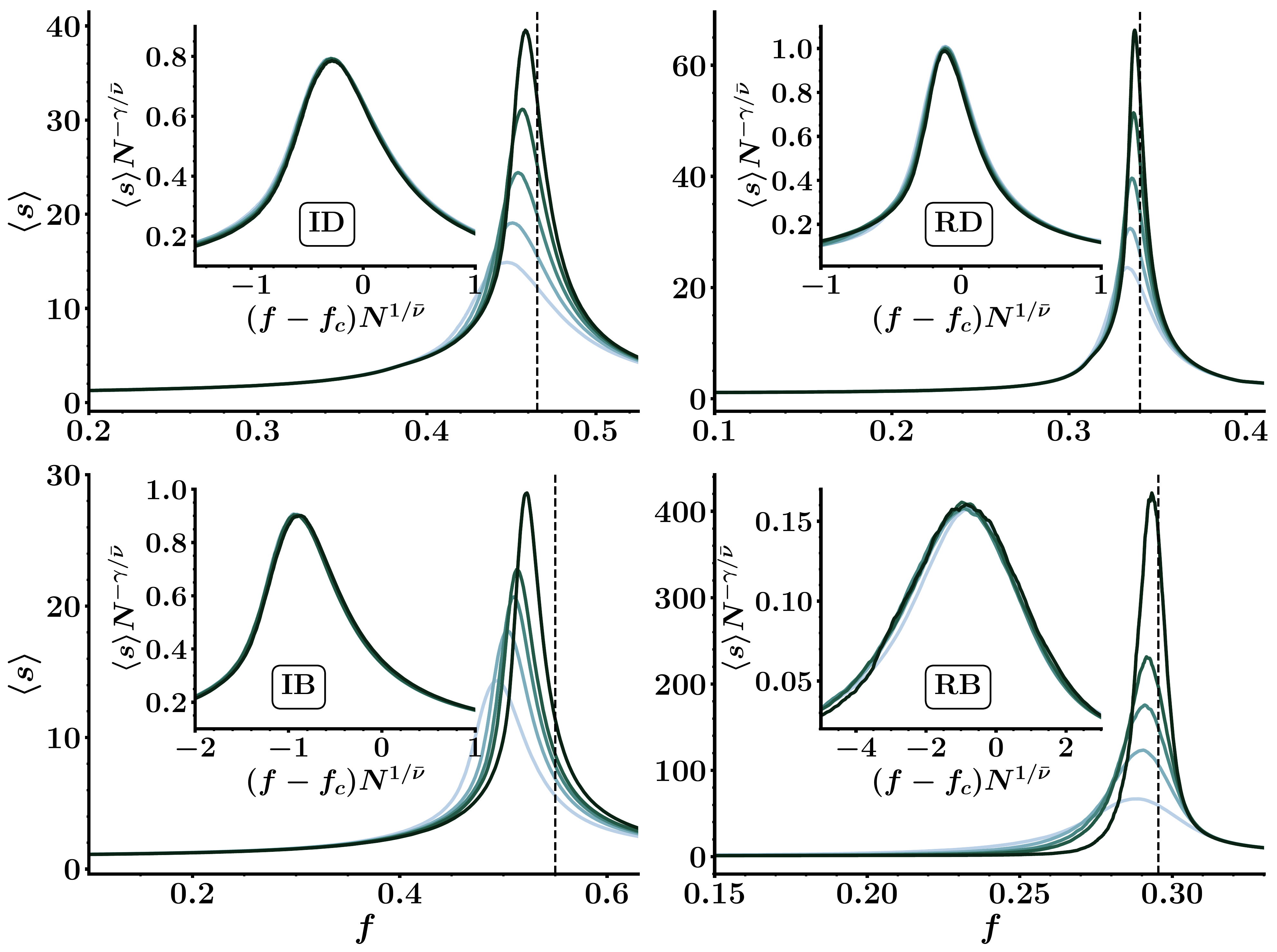}
    \caption{Susceptibility $\langle s \rangle$ as a function of the fraction of nodes removed $f$ for the four attack strategies studied. As it happens with the size of the second cluster, this quantity also peaks near the percolation threshold. The insets show the collapse of the curves using the scaling ansatz given by Eq. \ref{eq:suscep-scaling}, using the parameters given in Table \ref{table}. The color code is the same as in Figure  \ref{fig:S1_scaling}.}
    \label{fig:meanS_scaling}
\end{figure}

In Figures \ref{fig:N2_scaling} and \ref{fig:meanS_scaling}, the size of the second largest cluster $S_2N$ and the susceptibility $\langle s \rangle$ are shown. These quantities exhibit a peak close to the percolation threshold, which increases in magnitude with the system size. In the same way as with the order parameter, the curves can be scaled using Eqs. \ref{eq:comp-scaling} and \ref{eq:suscep-scaling}. The corresponding insets show that the collapses are good, thus confirming the validity of the scaling assumptions.

We focus now on the estimation of the critical exponents. By evaluating Eq. \ref{eq:suscep-scaling} at $f=f_c$, we have $\langle s \rangle (f_c, N)\sim N^{\gamma/\bar{\nu}}$, so a log-log plot of $\langle s \rangle$ vs $N$ at the percolation threshold should give a straight line with slope $\gamma/\bar{\nu}$. Thus,  the ratio between these two exponents can be computed directly using a linear fit. The main drawback of this method is that the percolation threshold must be known beforehand. As we only have an estimation for $f_c$, this method will propagate the uncertainty associated with that estimation. To avoid this, instead of computing the average cluster size at the percolation threshold, we compute the value at the peak of this measure performing the scaling using these values. We recall that, for sufficiently large system sizes, the scaling of the peaks is the same as the scaling at the percolation threshold \cite{Ramirez2018StandardLattices}.

\begin{figure}[h]
    \centering
    \includegraphics[scale=0.28]{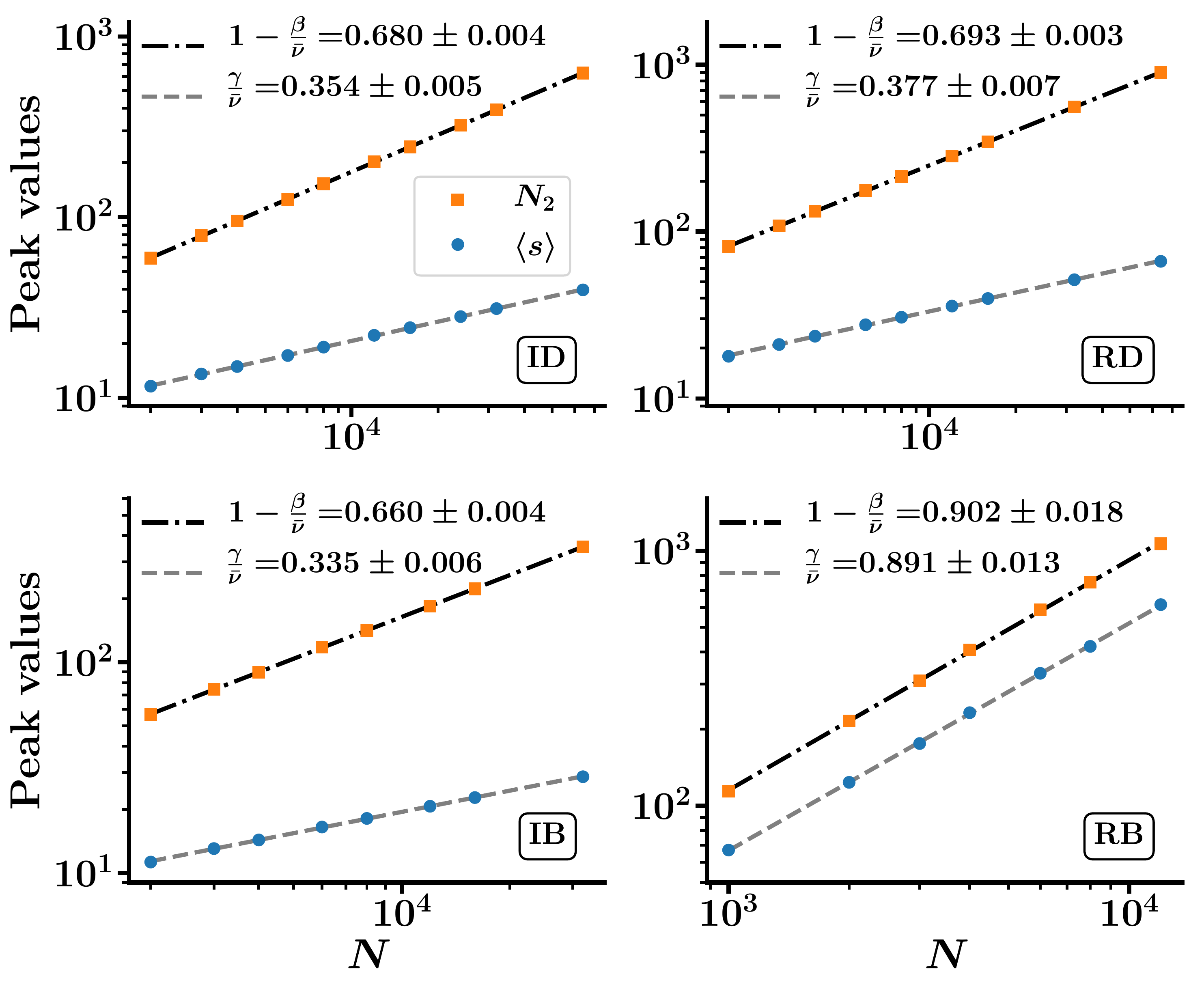}
    \caption{Scaling of the peaks of $S_2N$ and $\langle s \rangle$ for the four attack strategies. The markers represent the value at the peak, after averaging over $2\times 10^4$ simulations (ID, RD, and IB) and $10^3$ simulations (RB). Dashed lines correspond to a linear fit of the points using least-squares on a log-log scale.}
    \label{fig:peaks}
\end{figure}

In Figure \ref{fig:peaks}, we show the corresponding scaling for each of the four attack strategies implemented. In all cases, the linear relation in the log-log scale is very clear. The estimated ratios between the critical exponents are shown in the figure and summarized in Table \ref{table}. In a similar way, we note that the size of the second largest cluster $S_2N$ also peaks near the transition. According to Eq. \ref{eq:comp-scaling}, this quantity scales as $S_2N \sim N^{1-\beta/\bar{\nu}}$ near the critical point, so the ratio $1-\beta/\bar{\nu}$ can be inferred from the scaling of such peaks. Figure \ref{fig:peaks} also shows the values of the peaks and the corresponding linear fit. The estimated ratios are summarized, as before, in Table \ref{table}.

As a consistency check, we note that the estimated exponents satisfy Eq. \ref{eq:gamma_beta_nu}. The values obtained are $0.99(2)$ for ID, $0.99(2)$ for RD, $1.02(2)$ for IB and  $1.09(4)$ for RB.

\begin{figure}[h]
    \centering
    \includegraphics[scale=0.28]{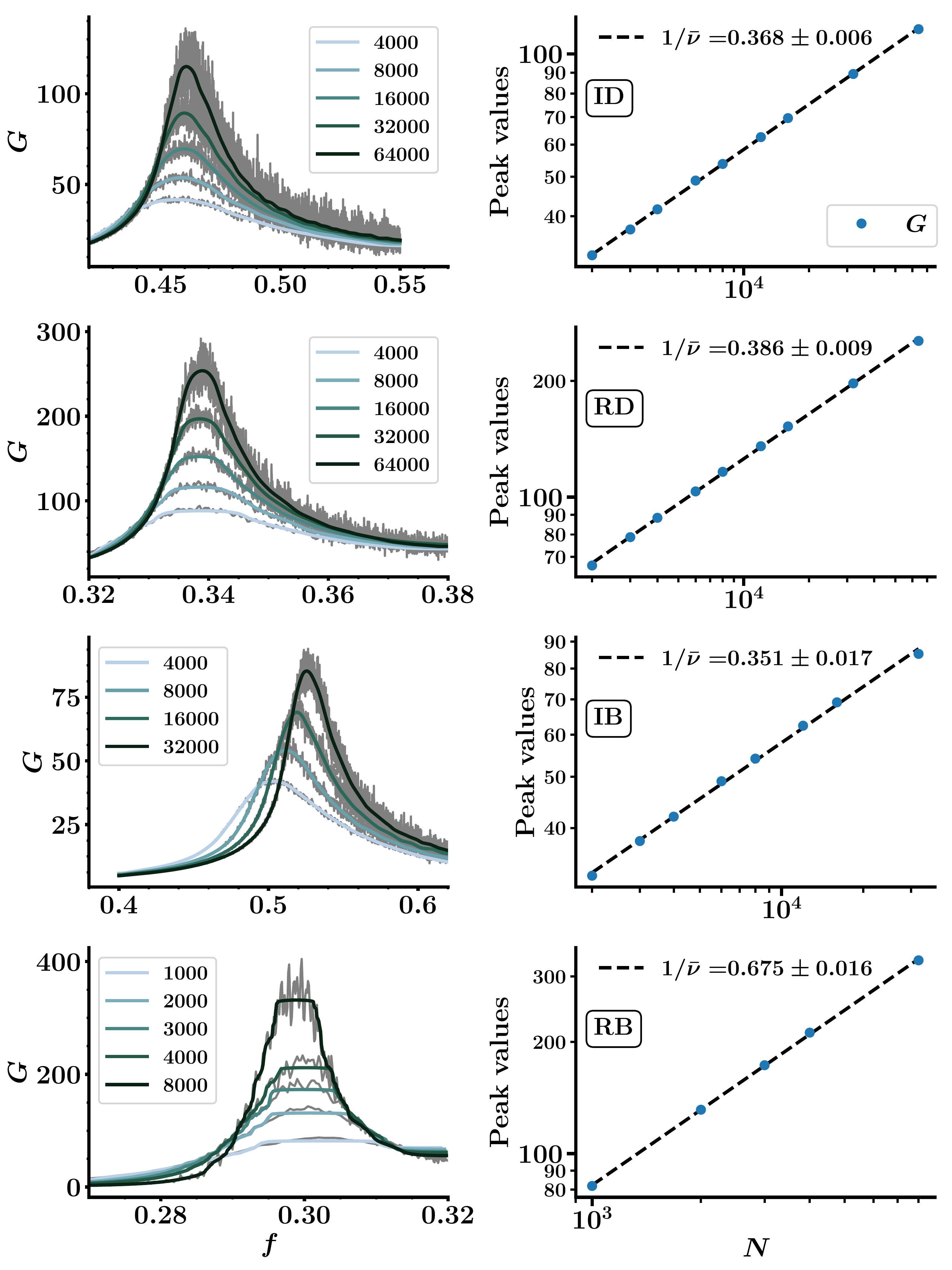}
    \caption{(Left panels) Derivative of the logarithm of the order parameter as a function of the fraction of nodes removed. The grey curves correspond to a five-point numerical derivative and the colored curves to the derivatives computed using the regularization method  described in \cite{Chartrand2011NumericalData}. Both methods were applied over averages using $2\times 10^4$ simulations (ID, RD, and IB) and $10^3$ simulations (RB). (Right panels) Scaling of the peaks of the curves in the left. Dashed lines correspond linear fits using least-squares. The values obtained for the critical exponent of the correlation length $\bar{\nu}$ are summarized in Table \ref{table}.}
    \label{fig:scaling_nu}
\end{figure}

To get the full characterization of each transition, it remains to compute the exponent $\bar{\nu}$. In order to do this, we define $G(f, N) = -\partial \log S_1(f, N)/\partial f$ and use Equation \ref{eq:comp-scaling} for $i=1$, from where we have 

\begin{align} \label{eq:deriv_log_Si}
    G(f, N) &\sim  -\dfrac{d \log \tilde{S}_1\left[ (f-f_c) N^{1/\bar{\nu}} \right]}{d f}  \nonumber \\
    &\sim N^{1/\bar{\nu}} \tilde{G}\left[(f-f_c) N^{1/\bar{\nu}}\right],
\end{align}
where $\tilde{G}(x) = -\tilde{S}_1'(x)/\tilde{S}_1(x)$. Then, the function $G$ has a similar scaling than the order parameter and the susceptibility. As the order parameter has an inflexion point close to the percolation threshold, then $G$ has a peak at that point. In the same way as it happens with the susceptibility and the second cluster, it is expected that the peaks scale as a power-law, this time, with an associated exponent $1/\bar{\nu}$. Then, we can perform a similar analysis as before and plot $G$ versus $N$, where we should see a linear relation in a log-log scale. This approach comes with the following caveat. In general, taking the numerical derivative of a noisy signal tends to amplify the noise. Our case is not an exception, as it can be seen in Figure \ref{fig:scaling_nu}. The grey curves correspond to the numerical derivative computed using five-point finite differences over the average of $2\times 10^4$ realizations  (ID, RD and IB) and $10^3$ simulations (RB). We can see that the noise is amplified and that it increases with the system size. To overcome this problem, we employed a regularization method, described in \cite{Chartrand2011NumericalData}, with which smoother curves can be obtained (colored curves). The right panels of Figure \ref{fig:scaling_nu} show the scaling of the peaks, computed from the regularized derivative. As it can be seen from the linear regressions, the scaling hypothesis is satisfied. The estimated values for the exponent of the correlation length are $\bar{\nu}^{\mathrm{(ID)}} = 2.72(5)$, $\bar{\nu}^{\mathrm{(RD)}} = 2.59(7)$, $\bar{\nu}^{\mathrm{(IB)}} = 2.8(2)$ and $\bar{\nu}^{\mathrm{(RB)}} = 1.50(6)$.

\subsection{Cluster size distribution}

\begin{figure}
    \centering
    \includegraphics[scale=0.28]{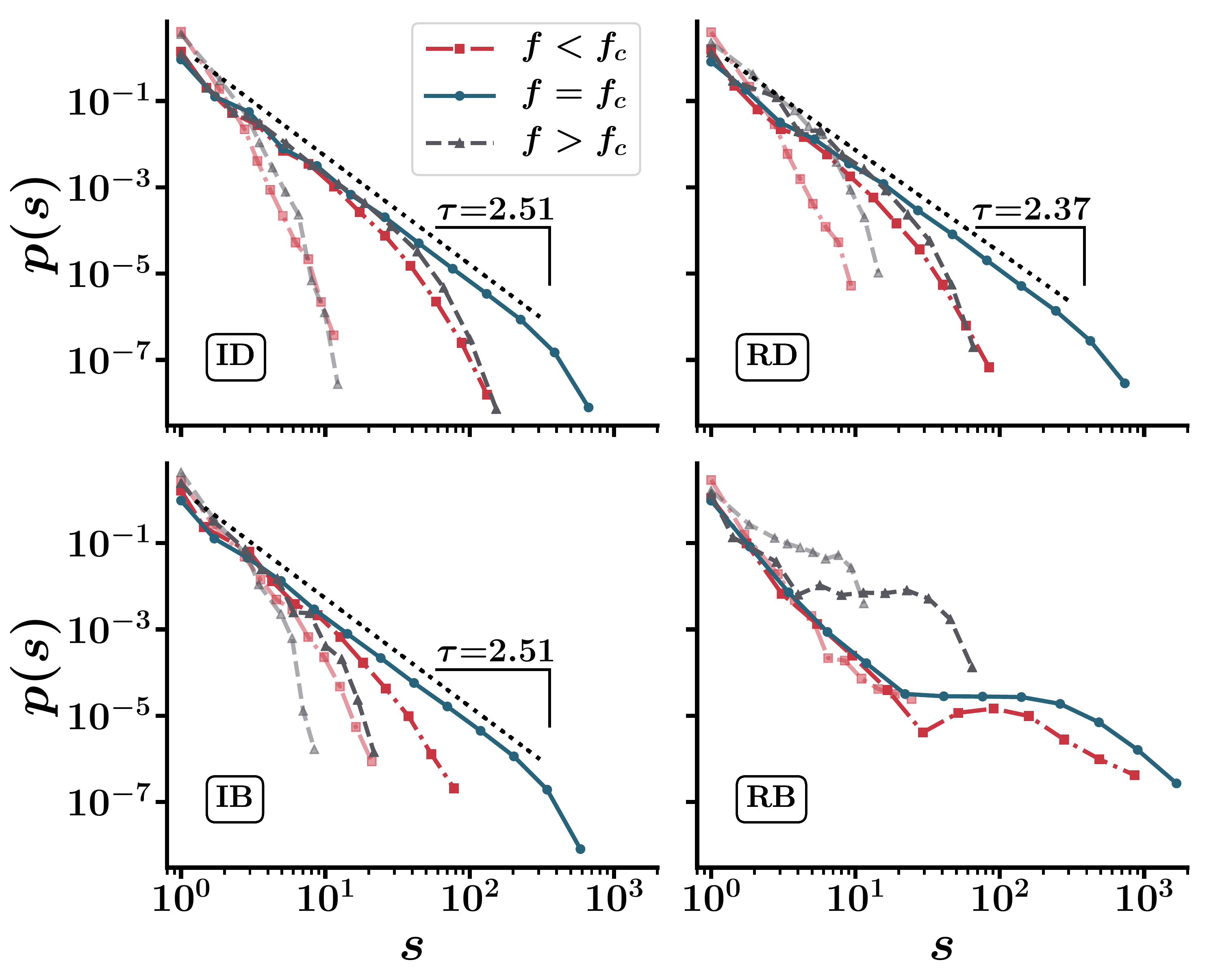}
    \caption{Cluster size distribution $p(s) = n(s) / \sum_s^{'} n(s)$ for each attack strategy. Dotted dashed red curves correspond to the sub-critical region $f < f_c$, dashed grey curves to the supper-critical region $f> f_c$ and solid blue lines to the critical value $f=f_c$. For ID, RD, and IB, the plots are consistent with equation \ref{eq:ns}, showing a power-law distribution at the critical point (with a gentle decay at the tail due to finite-size effects) and an exponential decay for other values of $f$. RB deviates from the standard behavior, showing a hump for middle values of $s$. The dotted black curves are power-laws fitted from the binned data, and their slopes are summarized in Table \ref{table}. The distributions were computed averaging $10^4$ networks for ID, RD, and IB and $10^2$ networks for RB. $N=16000$, $\langle k \rangle = 5$.}
    \label{fig:cluster_size}
\end{figure}

As it was previously explained, second-order percolation transitions exhibit a power-law cluster size distribution at the critical point given by Eq. \ref{eq:ns}. In Figure \ref{fig:cluster_size}, we show that this is indeed the case for the two degree-based attacks and the initial betweenness attack. The exponents of the respective power-laws---which were measured directly from $n(s)$ using a linear fit in logarithmic scale---are in agreement with the scaling relation given by Eq.  \ref{eq:scaling_tau} and are consistent, considering uncertainties, with the value $\tau = 2.5$ correspondent to standard percolation (see Table~\ref{table}). 
The case of recalculated betweenness deserves special consideration since it departs from the mean-field universality class as we point it out below. 
Although a power-law decaying can be seen for small cluster sizes, the distribution shows a hump at higher values departing from the expected behavior. 
In a similar manner to what happens with the abrupt drop in the order parameter near the transition, this behavior could be indicating 
a first-order phase transition. 
Nevertheless, it is worth noting that similar effects have been observed in other continuous transitions in the context of explosive percolation models \cite{DSouza2015AnomalousPercolation, Chen2012DerivingPercolation}. 
Here we argue that the hump is due to a finite-size transient effect and that it must disappear for larger system sizes. 
Using a heuristic argument similar to that of Ref.~\cite{DSouza2015AnomalousPercolation}, we can estimate a crossover size $N^*$, where the system becomes large enough so that realizations converge to the asymptotic limiting behavior. 
Let $\Delta S_{\mathrm{max}}$ be the greatest jump for the order parameter after removing a node in a single realization. 
The variation in the control parameter $f$ in this single step is $\Delta f = 1/N$. 
Assuming that this jump occurs at $f_c$ and using the scaling of the order parameter, we can roughly state that $\Delta S_{\mathrm{max}} \sim \Delta f^{-\beta} = N^{\beta}$. 
Now, we define $N^{*}$ as the system size for which the greatest jump in the giant component is about ten percent. 
Thus, $N^{*} \sim 10^{1/\beta}$. For the RB attack, $\beta \sim 0.15$ yielding $N^* \sim 10^6$. 
As the results presented in Figure \ref{fig:cluster_size} correspond to $N=16000$, we are still under the crossover size, which might explain the deviation from the power-law.  

\section{Discussion}\label{sec:Discussion}

Table \ref{table} summarizes the main results of this work, providing a characterization of the percolation transitions produced by the four attack strategies studied. From the perspective of the network dismantling problem, the relevant magnitude is the percolation threshold, which quantifies the strength of each attack. Our results not only confirm that the recomputed versions of the attacks perform better than their initial counterparts, which has been previously shown in the literature \cite{Holme2002AttackNetworks,Iyer2013AttackNetworks,Wandelt2018ANetwork}, but also allows us to quantify the amount of improvement that can be achieved by recomputing node centrality at each removal step.  If we compare the two degree-based attacks, the difference between their percolation thresholds is around $ \sim 0.12$. For the betweenness-based attacks, the improvement is $\sim 0.26$. As we see, the difference is greater in the latter case and the reason of this can be attributed to the global nature of the betweenness centrality, in contrast to the locality of the degree. 

From the point of view of critical phenomena, the critical exponents are the most relevant measures as they determine the universality class of the transition. 
Our results show that the two initial attacks have exponents that are consistent or close to the ones corresponding to random percolation. 
Besides, the scaling relations \ref{eq:scaling_tau} and \ref{eq:gamma_beta_nu} are satisfied within the uncertainty. The case of recalculated degree is similar in the sense that it also satisfies the scaling relations, but seems to differ in some of the exponents. 
The significance of the difference is not clear, however,  and in consequence, it is unclear if the attack belongs or not to the same universality class than the previous cases.
We note that in a previous work by Norrenbrock, et al.  \cite{Norrenbrock2016FragmentationAttacks}, the authors claim that a RD attack over two-dimensional proximity graphs has the same exponents as random percolation on a square lattice. 

Lastly, the recalculated version of the betweenness-based attack is qualitatively different from the rest of the attacks. The critical exponents are different from the mean-field values and the component size distribution does not exhibit a power-law decay for the system sizes studied. For its characteristics, this transition could be included in the framework of explosive percolation transitions \cite{Boccaletti2016ExplosiveSynchronization,DSouza2019ExplosiveNetworks}. Explosive transitions can be either continuous, with a steep derivative of the order parameter near the percolation threshold, or discontinuous, depending on the underlying process. Given that these transitions are usually characterized by a large crossover size, the order of the transition can be hard to determine. As it has been extensively discussed in recent reviews \cite{DSouza2015AnomalousPercolation,DSouza2019ExplosiveNetworks}, in some cases the transition seems continuous for finite-size systems but becomes discontinuous for large enough systems. In other cases, the opposite occurs. Moreover, there are transitions where both discontinuity and criticality coexist. Based on our results, we can safely say that RB has critical behavior and that it does not belong to the random percolation universality class. On the other hand, 
we cannot ensure the order of the transition from our methods, unless larger systems are studied, which seems unlikely in the short time given the computational complexity associated with the computation of betweenness.


\section{Conclusions}\label{sec:Conclusions}

We have studied the percolation transitions induced by four dismantling strategies based on centrality measures over Erd\H{o}s-R\'enyi networks. By performing a systematic finite-size scaling analysis, we have obtained both the percolation thresholds and the critical exponents that characterize the universality class of the transitions. By computing the percolation thresholds, we were able to verify and quantify the intuitive idea that the attack strategies become more effective when node centrality is updated after each removal step. In particular, we show that keeping updated information of node centrality can even modify qualitatively the percolation process, changing its universality class. 

From a dismantling point of view, recalculated betweenness is the most efficient attack, as it is the one exhibiting the lowest percolation threshold. In fact, its performance is comparable to the most effective methods to dismantle networks~\cite{Braunstein2016NetworkDismantling, Morone2015InfluencePercolation,Wandelt2018ANetwork}. Also, the critical exponents of the percolation process associated with this attack are far from trivial, and resemble the behavior observed in explosive percolation transitions~\cite{Achlioptas2009ExplosiveIn, Fortunato2011ExplosiveGraphs}.  
At variance with the degree-based attacks where the order parameter gradually decays towards zero, the dismantling with recalculated betweenness proceeds more silently, giving a misleading picture of integrity even at the edge of a catastrophic failure. If we think of infrastructures such as power grids, road networks or the Internet, it is reasonable to conceive heavy loaded nodes as the most prone to failure, so RB-like damages are possible not only as a targeted attack but as a failure.
Other authors have studied the vulnerability of these systems in terms of cascading failures using as a proxy for the loads 
the betweenness of the nodes~\cite{Kornbluth2018NetworkAttacks, Buldyrev2010}. 
From a novel perspective, our work suggests a new direction in which networked systems can be assessed in the search of critical vulnerabilities. 

\begin{acknowledgments}
This work was partially supported by grants from CONICET (PIP 112 20150 10028), FonCyT (PICT-2017-0973), SeCyT–UNC (Argentina) and MinCyT C\'ordoba (PID PGC 2018). In this work, we used Mendieta Cluster
from CCAD-UNC, which is part of SNCAD-MinCyT, Argentina. The authors thank Andr\'es Chacoma and Sergio Cannas for the useful discussions.
\end{acknowledgments}


%

\end{document}